%% file: ckm2012_jaeger.tex
\documentclass[12pt]{article}
\usepackage{epsfig}

\input econfmacros.tex
\def\Title#1{\begin{center} {\Large {\bf #1} } \end{center}}

\newcommand{\eq}[1]{{(\ref{#1})}}

\begin{document}

\Title{Theory of charmless hadronic $B$-decays}

\bigskip\bigskip

\begin{raggedright}  

{\it Sebastian J\"ager\index{J\"ager, S.}\\
Department of Physics and Astronomy\\
University of Sussex\\
Brighton BN1 9QH, United Kingdom
}
\bigskip\bigskip
\end{raggedright}

\begin{abstract}
I summarize results and performance of the dynamical theory of
charmless hadronic $B$ decays, based on QCD factorization in the heavy
quark limit. On the theoretical side, a number of NNLO ($\alpha_s^2$)
amplitudes are now available, all showing a well-behaved perturbative
expansion. The large observed branching fraction
in $\bar B^0 \to \pi^0 \pi^0$ remains a challenge, implying either
 a large inverse moment of the $B$-meson wave function or a sizable
power correction (or unexpected new physics). $B$-factory/Belle2 analyses
of $B \to \gamma \ell \nu$ may shed light on this.
On the other hand, the new Belle and LHCb measurements of $A_{\rm
  CP}(\pi^+\pi^-)$ bring the experimental result closer to
QCDF predictions, similar to what is found in $\bar B \to \pi \bar K$
decays, while $S_{\pi\pi}$ gives a competitive $\gamma$-determination.
I remind the reader that for vector-vector final states, a
theoretical treatment of the full set of helicity amplitudes has
existed and met with some success since the $B$-factory era, and applies
equally to LHCb measurements of e.g.\ $B_s \to \phi \phi$.
\end{abstract}

\vfill

\begin{center}
Proceedings of CKM 2012, the 7th International Workshop on the CKM Unitarity Triangle, University of Cincinnati, USA, 28 September -- 2 October 2012 
\end{center}

\section{Introduction}
Charmless hadronic $B$-decays have been at the center of experimental
and theoretical interest for many years. The reasons are threefold:
(i) there are a large number of measurements, with over 100 final states
when considering light pseudoscalars and vectors alone;
(ii) they are sensitive both to CKM elements and, being
short-distance-dominated rare processes, to possible new heavy
particles;
(iii) they are conceptually interesting as they involve an intricate
interplay of the three different Standard-Model (SM) interactions
and the hierarchy of energy scales
 $ M_W$, $m_b$, $\Lambda_{\rm QCD}$.

Concretely, any weak $B$-decay into two
charmless hadrons has an amplitude
\begin{equation}  \label{eq:charmless}
  {\cal A}(\bar B \to M_1 M_2) =
     e^{-i \gamma} |V_{uD} V_{ub}| T_{M_1 M_2}
     + |V_{cD} V_{cb} |P_{M_1 M_2}
     + {\cal A}_{\rm NP} ,
\end{equation}
where $D=d$ or $s$, the ``tree'' $T_{M_1 M_2}$ and ``penguin'' $P_{M_1 M_2}$
are CP-even ``strong'' amplitudes, comprising hadronic matrix elements
of the weak Hamiltonian ${\cal H} = \sum_i C_i Q_i$, and
${\cal A}_{\rm NP}$ denotes a possible beyond-SM (BSM) contribution.
Extracting CKM information or identifying a BSM contribution from the
data requires knowledge about the strong amplitudes  $T_{M_1 M_2}$
and $P_{M_1  M_2}$.

Isospin symmetry may be used to group together classes of
decays (such as all $B \to K \pi$ decays),  which may involve
up to two tree amplitudes (colour-allowed and colour-suppressed),
a QCD-penguin amplitude, electroweak penguin amplitudes, and so
on. These ``topological'' amplitudes (first row of Table
\ref{tab:topo-hierarchies})
can be visualized as weak-interaction diagrams.
\begin{table}[tb]
\centering
\caption{Topological amplitudes. First row: notation prevailing in the
data-driven/$SU(3)$-based literature. Second row: notation used
in the context of QCD factorization (originating from naive
factorization). Further rows: scaling in powers of
  the Cabibbo angle $\lambda$, in $1/N_c$, and in $\Lambda_{\rm
  QCD}/m_b$. Some multiply suppressed amplitudes
are omitted.
\label{tab:topo-hierarchies}}
\begin{tabular}{c|cccccccccc}
traditional name & $T$ & $C$ & $P_{ut}$ & $P_{ct}$ &
$P_{EW}$ & $P_{EW}^{\rm C}$ &
$(P_{ct})$ & $PA$ & $E$ & $A$ \\
notation of \cite{Beneke:1999br} & $a_1$ & $a_2$ & $\alpha_4^u$ &
$\alpha_4^c$ & $\alpha_{3EW}$ & $\alpha_{4EW}$ &
$\beta_3^c$ & $\beta_4^c$ & $\beta_1$ & $\beta_2$ \\

\hline
Cabibbo $(b\to d)$ & \multicolumn{10}{c}{all amplitudes are
${\cal O}(\lambda^3)$ } \\
Cabibbo $(b\to s)$ & $\lambda^4$ & $\lambda^4$ &
$\lambda^4$ & $\lambda^2$ & $\lambda^2$ & $\lambda^2$ & $\lambda^2$ &
$\lambda^2$ & $\lambda^4$ & $\lambda^4$  \\
$1/N$ & $1$ & $\frac{1}{N}$ & $\frac{1}{N}$  & $\frac{1}{N}$ &
$1$ & $\frac{1}{N}$ & $\frac{1}{N}$ & $\frac{1}{N}$ & $\frac{1}{N}$ & $1$ \\ 
$\Lambda/m_b$ & $1$ & $1$ & $1$ & $1$ & $1$ & $1$ & $\Lambda/m_b$ & $\Lambda/m_b$ & $\Lambda/m_b$ & $\Lambda/m_b$ \\
\end{tabular}
\end{table}
Existing theoretical treatments either
 eliminate the strong amplitudes
    by simultaneously considering $b \to d$ and $b \to s$ transitions
    and invoking $SU(3)_F$ symmetries
    \cite{Zeppenfeld:1980ex,Gronau:1994rj} (at the expense of a smaller
number of BSM-sensitive observables), or attempt to calculate some
    strong amplitudes,
or some combination thereof.
Of the computational approaches, those that achieve some degree of
model-independence are based on an
expansion in $\Lambda/m_b$: QCD factorization (QCDF)
\cite{Beneke:1999br,Beneke:2000ry,Beneke:2001ev}
and its effective-field theory formulation in SCET
\cite{Bauer:2000yr,Bauer:2004tj,Beneke:2005vv}, and the still more
ambitious, but also more model-dependent,  pQCD approach
\cite{Keum:2000ph,Keum:2000wi}. 

\section{Status of QCD factorization}
\paragraph{Structure}
The QCD factorization approach \cite{Beneke:1999br,Beneke:2000ry} is
based on the collinear factorization of the hadronic matrix elements
$\langle M_1 M_2 | Q_i | \bar B \rangle$  in the heavy-quark limit. To leading
power in $\Lambda/m_b$, the collinear and soft dynamics is contained
in $B\to \mbox{light hadron}$ form factors and light-cone distribution
amplitudes (LCDA) for the initial- and final-state mesons, convoluted with
perturbative hard-scattering kernels. Schematically,
\begin{eqnarray}
  F_{B M_1} f_{M_2}\, a &=& F_{B M_1} f_{M_2} \int t^{\rm I}(u) \phi_{M_2}(u) du
\nonumber \\
   & & + f_{B} f_{M_1} f_{M_2} \int t^{\rm II}(u, v) \phi_{M_2}(v)
   \phi_{M_1}(u)\, du \,dv,   \label{eq:bbns}
\end{eqnarray}
where now $a$ denotes any of the amplitudes
in Table \ref{tab:topo-hierarchies},
$F_{B M_1}$ is a form factor and $f_B$, $f_{M_1}$, $f_{M_2}$ are decay
constants.\footnote{By convention,
a product of a form factor and a decay
  constant is factored out of the $a_i$ / $\alpha_i$ amplitudes,
  but not in \eq{eq:charmless} (nor in most $SU(3)$-based analyses).
  Accordingly the normalization conventions differ between the first
  and second rows of Table \ref{tab:topo-hierarchies}.}
 The kernels $t^{\rm I}$ and $t^{\rm II}$ are
hard-scattering kernels and are perturbatively calculable as power
series in $\alpha_s$. The structure \eq{eq:bbns} holds at the leading
power in $\Lambda/m_b$ and is unambigous, free from scale or scheme
dependences, etc. Some important terms at the
first subleading power also factorize, but others do not
(an attempt to factorize them leads to endpoint-divergent convolutions).

\begin{figure}[t]
  \vspace*{-4cm}
    \includegraphics[scale=0.5]{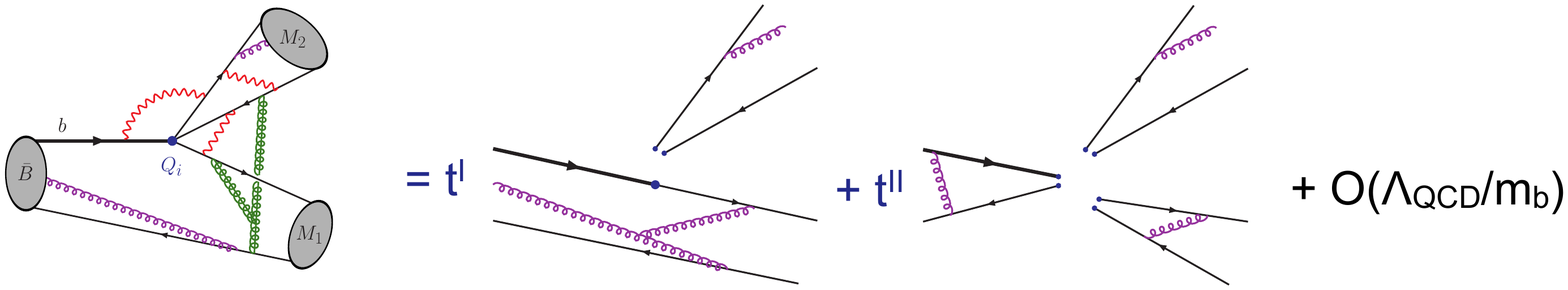}
\vspace*{-4cm}
 \caption{Matching calculation for the hard-scattering kernels. The
   red lines are hard (virtuality $\sim m_b^2$), the green lines
   hard-collinear (virtuality $\sim m_b \Lambda$). Gluon exchanges of
   lower virtualities are reproduced by the effective-theory matrix
   elements, which, for hadronic states, define form factors and
   light-cone distribution amplitudes.
\label{fig:matching} }
\end{figure}

The hard-scattering kernels are computed by considering appropriate
partonic states with the quantum numbers of the initial and final
mesons, consisting of soft partons for the $B$ meson and collinear
ones for the two final-state mesons (Figure \ref{fig:matching}).  %
The structure of \eq{eq:bbns} emerges most transparently within
soft-collinear effective theory (SCET),
whereby the hard kernels become Wilson coefficients and the
form factors and light-cone distribution amplitudes become matrix
elements of operators in the effective theory; the full equivalence
(up to a change of operator basis) to the original, diagram-based formula
has been demonstrated in \cite{Beneke:2005vv,Beneke:2006mk}.
Any differences in practice arise from (independent)
approximations at the phenomenological stage, such as neglect of
higher orders or the treatment of power corrections. (In this
context see also \cite{Beneke:2009az}.)

At the leading power, only
two-particle partonic states need to be considered. The $B$-meson
lines can either annihilate via the weak Hamiltonian (these terms can be
shown to be power-suppressed), or the valence light quark in the
$B$ meson can
form a spectator line with one of the lines representing $M_1$.
The ``hard-spectator-scattering'' terms on the second line of \eq{eq:bbns}
involve all diagrams that include a hard gluon exchange involving the
spectator line (green lines in Figure \ref{fig:matching}).
 However, there is a second leading-power contribution
from the end-point region where the spectator quark enters $M_1$ as a
soft parton. These diagrams are responsible for the first line in \eq{eq:bbns}.
As a consequence, at zeroth order in $\alpha_s$ and in the
heavy-quark limit, ``naive factorization'' is obtained.

The pQCD approach \cite{Keum:2000ph,Keum:2000wi}
aims to also factorize the form factors,
introducing some new conceptual issues
%(unsolved, in our opinion) 
and parametric dependences.
We refer to the original literature for more details.

\paragraph{Status of the perturbative kernels} Over the last 7 years a
number of substantial works have pushed the precision from NLO
($\alpha_s$) order (at which nontrivial factorization of infrared
physics first occurs) to NNLO ($\alpha_s^2$),
beginning with spectator scattering for the leading-power tree
\cite{Beneke:2005vv,Kivel:2006xc,Pilipp:2007mg}
and (QCD and QED) penguin \cite{Beneke:2006mk,Jain:2007dy} amplitudes,
followed by the form factor terms for
the trees \cite{Bell:2007tv,Bell:2009nk,Beneke:2009ek}.
In all cases, a well-behaved perturbation series is obtained, and the
structure of \eq{eq:bbns} holds, i.e.\ infrared physics factorizes as
expected. While no formal all-orders proof of factorization has been
published, in view of the high complexity and intricate
cancellations observed, this should be considered strong evidence
that factorization holds at higher orders.
The missing piece at NNLO at the leading power is the two-loop
form-factor correction to the penguin amplitudes.\footnote{A
partial calculation exists, comprising the chromomagnetic-operator
contribution \cite{Kim:2011jm}.}

\paragraph{Power corrections}
In phenomenological applications certain terms that are formally
$\Lambda/m_b$-suppressed cannot be neglected. First, there is the
so-called ``scalar-penguin'' amplitude $a_6^c$. This forms part
of the QCD penguin amplitude $\alpha_4^c$ and includes the hadronic
matrix element of the $(V-A) \times (V+A)$ QCD penguin operator $Q_6$,
as well as ``charming-penguin'' loop contributions. (The superscript
refers to the CKM structure $V_{cb} V_{cD}^*$ in \eq{eq:charmless}.)
For some of the final states that involve at least one pseudoscalar,
this contribution is ``chirally enhanced'' by a large normalisation factor.
E.g.\ for $M_2$ a pion, the
normalisation $r_\chi^\pi =  2\, m_\pi^2/(m_b (m_u + m_d))$
is formally power-suppressed but numerically
greater than one.
Fortunately the scalar penguin factorizes. At the moment, the NNLO
contributions, which might be phenomenologically relevant (see below),
are not known. A second, likely important power correction is the
annihilation amplitude $\beta_3^c$. (The possible importance of such a term,
which involves a large colour factor, was first pointed out in the
pQCD framework \cite{Keum:2000ph}.)
Note that ``annihilation'' refers to the
way the external lines are contracted with the weak Hamiltonian in the
matching onto SCET, and not to a topological annihilation amplitude:
Like $a_6^c$, $\beta_3^c$ forms part of the topological (QCD-)penguin
amplitude, which decomposes in the heavy-quark limit as
\begin{equation}  \label{eq:penganatomy}
  P_{ct} \propto  \hat \alpha_4^c \equiv \alpha_4^c + \beta_3^c
\equiv a_4^c \pm r_\chi^{M_2} a_6^c + \beta_3^c .
\end{equation}
The sign in front of $a_6^c$ depends on the spin of the final-state
meson $M_1$. $\beta_3^c$ (and other annihilation amplitudes) do
not factorize and need to
be modelled, usually according to the parameterization
given in \cite{Beneke:2001ev}.
Thirdly, there are contributions to the topological amplitudes which
arise at the level of higher-twist LCDA's for $M_1$ or multi-particle
LCDA's for the $B$ meson, which do not factorize.
These terms are (by convention) included in the amplitudes $a_{1,2,4}^{u,c}$;
in the SCET formulation, they involve power-suppressed parts
of SCET${}_{\rm I}$ matrix elements.
$a_4^c$ also  includes \textit{power-suppressed}  ``charming penguin''
corrections, for which however no enhancement mechanism has been
identified.

\paragraph{Electroweak amplitudes, singlets, vector-vector final states, etc}
The electroweak penguin operators in the weak Hamiltonian, together
with QED effects, generate further topological amplitudes which
factorize at the leading power. If
one or both of the final-state mesons are $SU(3)_F$ singlets, some
extra amplitudes appear. We refer to the original
literature, specifically \cite{Beneke:2003zv} (see also
\cite{Williamson:2006hb}), for a comprehensive
discussion. If both final-state particles are vector mesons, the
number of amplitudes triples. In this case, only the helicity-$0$
(``longitudinal'') amplitudes factorize. The other helicity amplitudes are
power-suppressed, but are numerically not negligible \cite{Kagan:2004uw}.

\section{Phenomenology}
Generically, QCDF implies that the predictions of naive factorization hold
up to corrections of order $\alpha_s$ or $\Lambda/m_b$, in particular
(i) direct CP asymmetries are typically suppressed, (ii) the corrections $\Delta
S_f$ to the sine-coefficients in time-dependent CP asymmetries, due to
sub-leading amplitudes, are given by naive factorization up to small
corrections. Moreover, pure-annihilation modes are power-suppressed.
All of this matches with the available data, with
some well-known qualifications.

The topological tree amplitudes are known to NNLO. The colour-allowed
and colour-suppressed trees $a_1$ and $a_2$ evaluate to \cite{Beneke:2009ek}
\begin{equation}
  a_1 = 1.000^{+0.029}_{-0.069} + (0.011^{+0.023}_{-0.050} ) \, i\, ,
\qquad
  a_2 = 0.240^{+0.217}_{-0.125} + (-0.077^{+0.115}_{-0.078}) \,i
\end{equation}
for a $B \to \pi \pi$ decay, where all errors have been combined in quadrature.
We see that the colour-allowed tree amplitude carries a very small
uncertainty and a tiny strong phase (and is very close to the
naive-factorisation result $C_1 + C_2/3 = 1.009$). On the other hand,
the colour-suppressed tree amplitude carries a large uncertainty. The
detailed anatomy of the NNLO central value including twist-3 power corrections is
\cite{Beneke:2009ek}
\begin{eqnarray}
a_2 &=& \Big\{
[0.220]_{\rm LO} + [-0.179 - 0.077 i]_{\rm NLO} + [-0.031 - 0.050 i]_{\rm NNLO}
\Big\}_{\rm FF} \nonumber \\
& & + \left[ \frac{r_{\rm sp}}{0.445} \right] \Big\{ [0.114]_{\rm NLO}
+ [0.049 + 0.051 i]_{\rm NNLO} + [0.067]_{\rm tw3} \Big\}_{\rm spec} .
\end{eqnarray}
Here, the two lines correspond to the two lines of \eq{eq:bbns}.
There is strong destructive interference between the LO term and
the higher-order form-factor corrections, which amplifies the relative
importance of spectator scattering and of power corrections. In addition,
the form-factor and spectator-scattering corrections interfere destructively
with each other.
The spectator scattering  suffers moreover from an uncertain
normalization
$r_{\rm sp} = (9 f_{M_1} \hat f_B)/(m_b \lambda_B f_+^{B\pi}(0))$.
This crucially depends on the first inverse moment $1/\lambda_B$ of
the relevant $B$-meson LCDA, which is poorly known. Experimental data
on $BR(\bar B\to \pi^0 \pi^0$) suggest a very large magnitude of
$a_2$; this may point to a small value of $\lambda_B$, or to an
underestimate of the twist-3 spectator-scattering power
correction.  Further $B$-factory measurements of $B \to \gamma
\ell \nu$, which is very sensitive to $\lambda_B$, may help with
resolving this; see \cite{Beneke:2011nf,Braun:2012kp} in this context.
Measurements with $\pi^0 \rho^0$ and $\rho^0 \rho^0$
show less drastic discrepancies; see also \cite{Bell:2009fm}.
The remaining amplitudes (QCD and EW penguins) are free from
such strong cancellations, and consequently carry more modest
uncertainties. They are currently known only to NLO.

If the aim is to use QCDF to search for BSM
effects, then its phenomenological usefulness must be first
judged against data on BSM-insensitive observables. This has been done
for modes dominated by tree and QCD penguin amplitudes in the SM, based on
the argument that these are in turn dominated by SM $W$-boson 
exchange at tree-level or within a charm loop. 
This class includes all $B \to \pi \pi$
observables as well as the direct CP asymmetry $A_{\rm CP}(\bar B \to\pi^+ K^-)$,
and related observables obtained by replacing one of the
final-state mesons by a vector ($\pi \to \rho$ or $K \to K^*$).
The case of $BR(\bar B \to \pi^0 \pi^0)$ has already been discussed.
\begin{figure}[t]
    \includegraphics[scale=0.2]{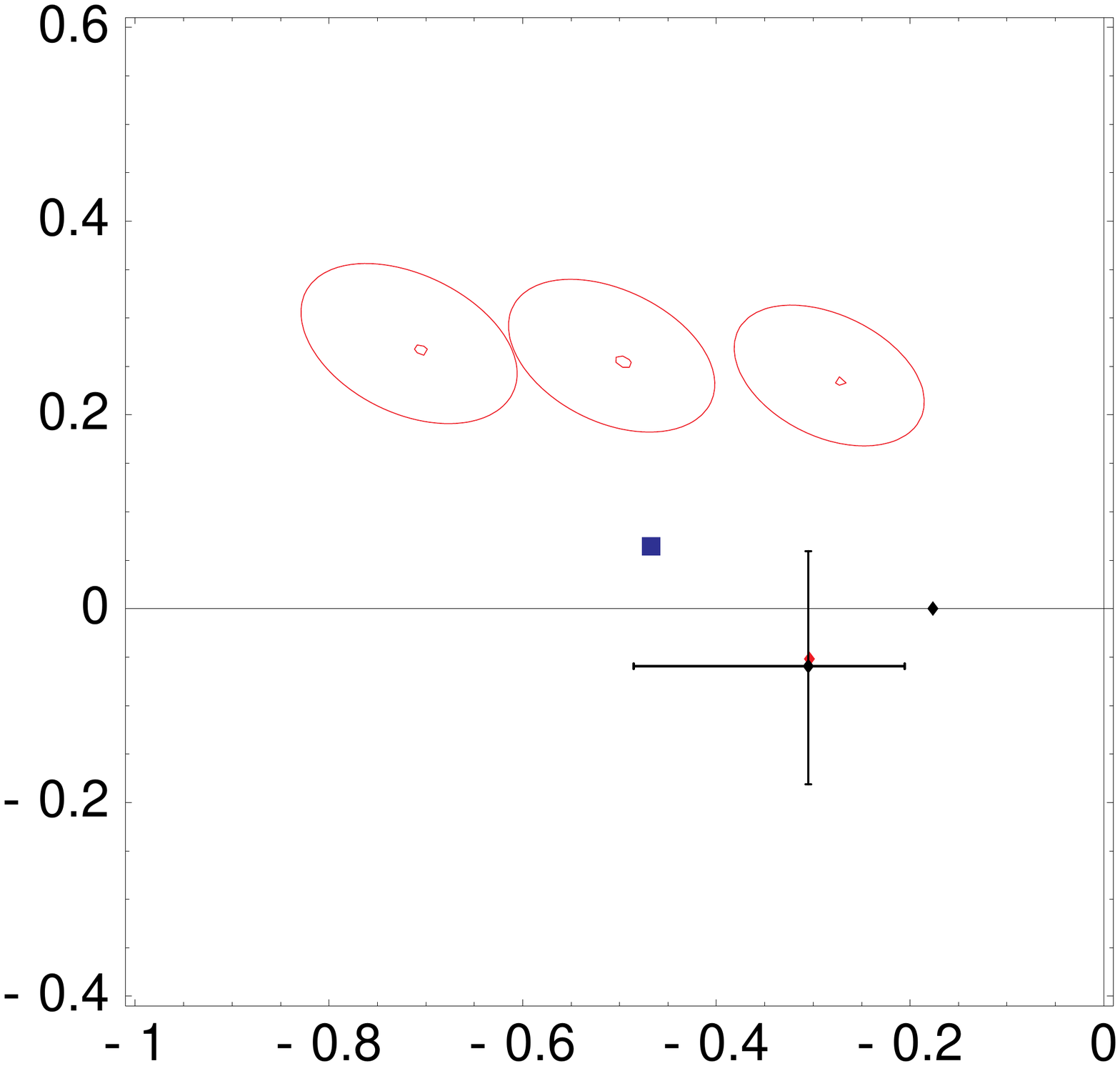}
\hspace*{0.4cm}\parbox{0.7\textwidth}{
\vspace*{-5.35cm}
    \includegraphics[scale=0.45]{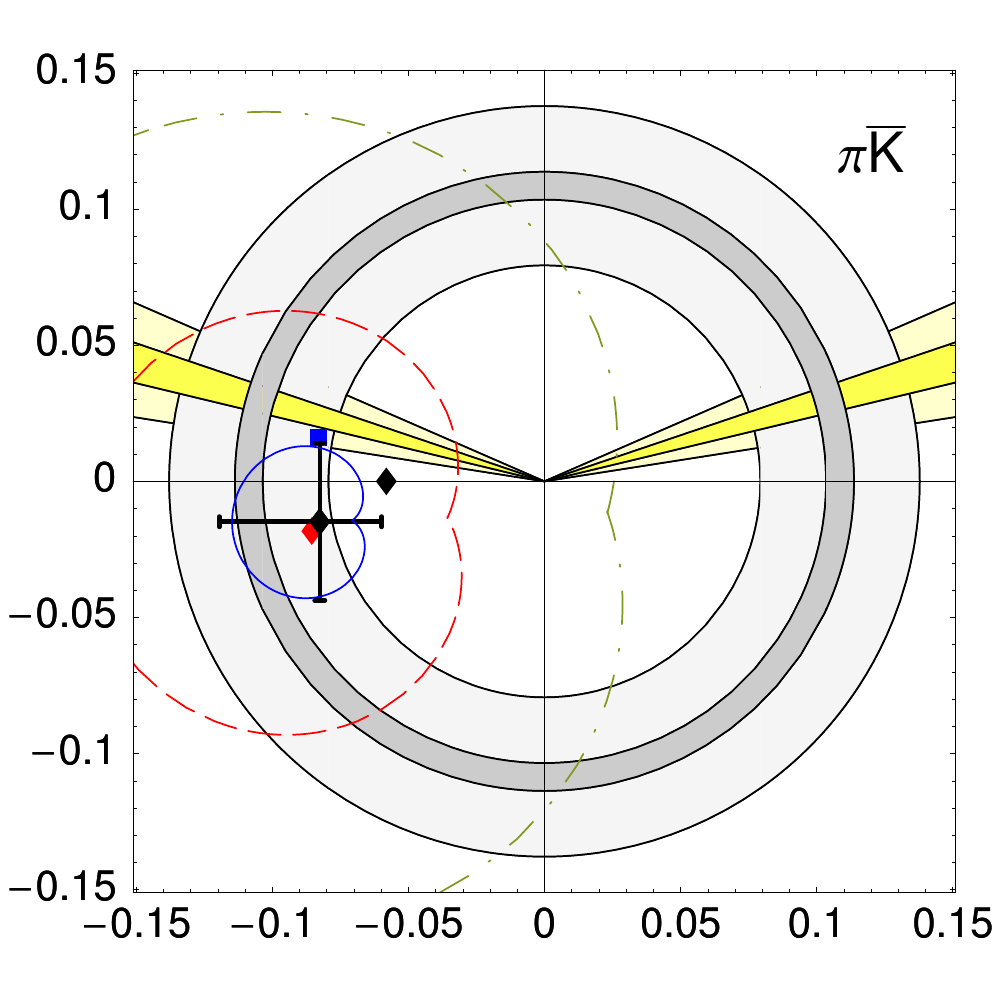}
\hspace*{0.3cm}
    \includegraphics[scale=0.45]{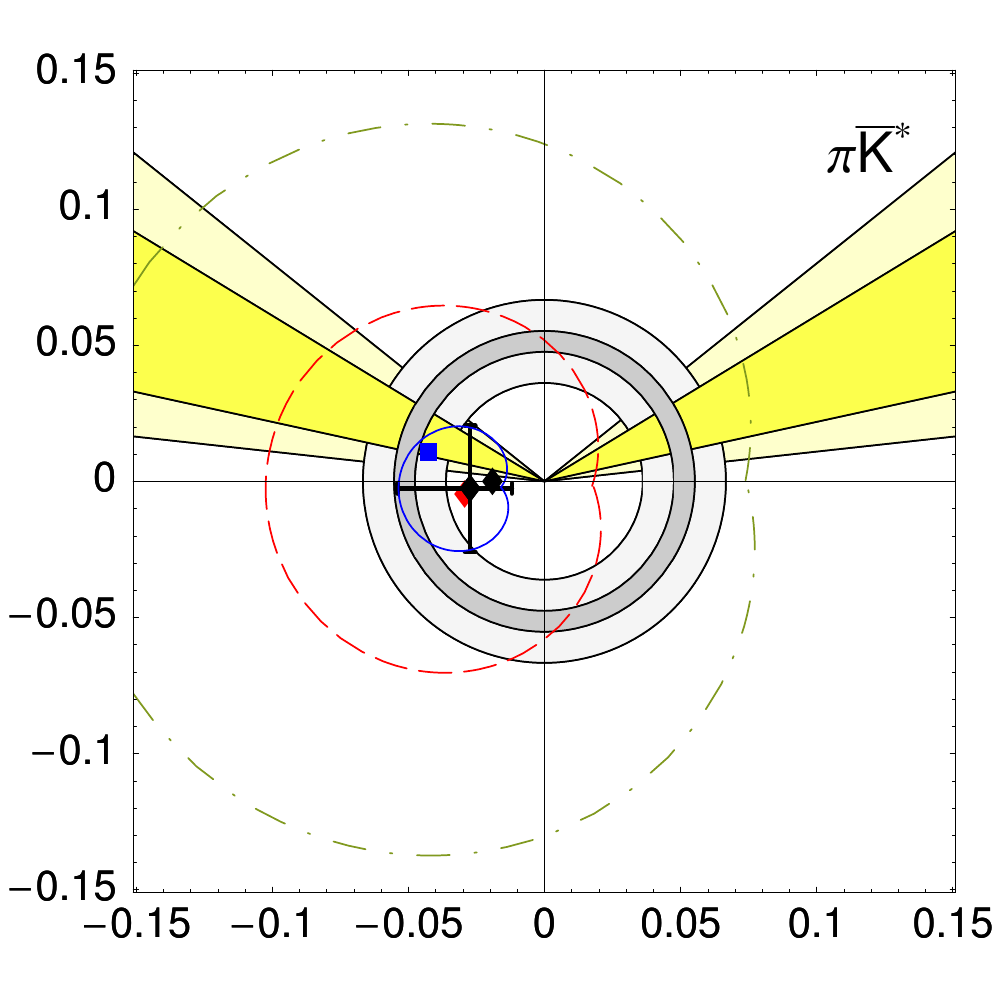}
}
 \caption{Penguin-to-tree ratios: fits to data compared to QCDF
predictions (points with error bars).
Left panel: $P_{\pi^+\pi^-}/T_{\pi^+\pi^-}$. The three ovals
correspond to (from right to left)
$\gamma=60^\circ, \gamma=70^\circ, \gamma=80^\circ$.
Middle panel: $\hat \alpha_4^c(\pi \bar K)/(a_1(\pi\pi)+a_2(\pi\pi))$,
right panel: $\hat \alpha_4^c(\pi \bar
K^*)/(a_1(\pi\pi)+a_2(\pi\pi))$. See text for details.
\label{fig:validation} }
\end{figure}
Figure \ref{fig:validation} (left) shows a complex penguin-to-tree
ratio that can be extracted from the time-dependent CP-asymmetry
in $\bar B \to \pi^+ \pi^-$, given the CKM angle $\gamma$. The
ellipses correspond to $1\sigma$ experimental errors, the cross
denotes the QCDF prediction with errors combined in quadrature,
and the blue square
corresponds to the parameter set ``G'' \cite{Beneke:2005vv} which
accomodates data on $\bar B \to \pi \pi$ and $\bar B \to \pi \bar K$ well.
The fitted imaginary part is weakly sensitive to $\gamma$ and opposite
in sign from expectations. The discrepancy is less
striking than in the past, however, due to the significantly reduced value of
$|C_{\pi\pi}|$ reported by Belle at this conference
\cite{Vanhoefer:2013rc,Adachi:2013mae},
and the small LHCb
result \cite{LHCbNote}. Note also that the theory prediction  may still
receive an important charm-loop correction
(proportional to the large Wilson coefficient $C_1$). In
 spectator scattering, such contributions first enter at
order $\alpha_s^2$ and are not known for $a_6$ yet.
The fitted real part is in agreement with theory
for $\gamma \sim 60^\circ-70^\circ$. Turning this around, one can say
that QCDF allows for a $\sim (5-7)^\circ$-precision
determination of $\gamma$ from $\bar
B \to \pi^+ \pi^-$ alone. The middle and right panels in
Fig.\ \ref{fig:validation} show
another penguin-to-tree ratio that can be extracted from $\bar B \to
\pi \pi$ and  $\bar B \to \pi \bar K$ (middle) and $\bar B \to \pi
\bar K^*$ data (see \cite{Beneke:2003zv} for details). The
fitted value corresponds to the intersection of the left wedge
and the circle. The blue, onion-like shape delimits an estimate of
the complex annihilation contribution $\beta_3^c$. In the case of
$\pi\bar K$, one observes a reasonable comparison between theory and
data within errors. This is quite a nontrivial check -- the fitted
result could a priori lie anywhere in the part of the complex plane
depicted (or even outside).
The discrepancy (mainly) in the imaginary part simply
reflects that there is no precise prediction for the direct CP
asymmetry $A_{\rm CP}(\pi^+ K^-)$, and is very similar (involving
$\hat \alpha_4^c$) to what is seen for $\bar B \to \pi^+ \pi^-$. Finally,
in the case of $\pi \bar K^*$ one has agreement between QCDF and data
within errors. The magnitude of the penguin amplitude $\hat
\alpha_4^c$ is significantly smaller than in the $\pi \bar K$
case. This can be understood based on the structure in
\eq{eq:penganatomy}: In the case of a vector $M_2 = \bar K^*$, the
scalar-penguin contribution $a_6^c$ is no longer chirally enhanced,
and effectively absent. This is a characteristic prediction of the
heavy-quark limit; I am not aware of any alternative theoretical
explanation.

It is also worth noting that several dedicated studies of
vector-vector final states exist, notably
\cite{Kagan:2004uw,Beneke:2006hg} dating from the $B$-factory
era. As said above, two out of three helicity (or equivalently
transversity) amplitudes are power-suppressed, but as pointed out
in \cite{Kagan:2004uw}, in the case of the penguin-amplitudes
this suppression is effective only for the positive-helicity amplitude
(in a $\bar B$ decay). Modelling power corrections in the usual way,
in the case of  $\bar B \to \phi K^*$, the full angular analyses
performed by Belle and Babar can be reproduced by QCDF \cite{Beneke:2006hg}.
Much of the theory carries over to
$B_s \to \phi \phi$, under study at LHCb \cite{Aaij:2012ud}.
The suppression of, for example, certain triple-product asymmetries is
a simple consequence of (and depends on!) the survival of the
suppression of the positive-helicity amplitude in the presence of QCD
effects. For a comprehensive study focusing on the (calculable)
longitudinal-polarization observables, see \cite{Bartsch:2008ps}, which
also contains references to ealier work on vector-vector final states in
the heavy-quark expansion.

Finally, a large number of studies of BSM effects employing QCD
factorization  exist, but it is not possible to do justice to them
in the limited space here. However, the discussion here should convey
that there are observables which remain NP-sensitive, after
uncertainties on strong amplitudes are taken into account.

\section{Conclusion}

The QCD factorization approach provides a consistent and unambigious
framework for computing hadronic $B$ decay amplitudes, including
strong phases, in the heavy-quark limit. The leading-power amplitudes
are partly available at NNLO, and further calculations by several
groups are underway. The same is true for the factorizable
``scalar-penguin'' power corrections. All results so far point to a
well-behaved perturbation expansion and no violations of factorization
have been found nor are any expected. At the phenomenological level,
upon modelling non-factorizable power corrections with a reasonable model
one obtains a consistent description of most data within the Standard
Model, within uncertainties. This includes many nontrivial
predictions. The main tension is with the $B \to \pi^0 \pi^0$ mode,
where new-physics contributions are generally not expected. This could
be explained through an underestimated power correction or a
large inverse moment of the $B$-meson LCDA. Future experimental data
may help.  The framework extends to vector-vector final states and can
accomodate $B$-factory data on $B \to \phi K^*$, and can make predictions
for LHCb measurements of the angular distribution in $B(B_s) \to V V$.

\end{document}

%% file: econfmacros.tex
\def\beq{\begin{equation}}
\def\eeq#1{\label{#1}\end{equation}}
\def\eeqn{\end{equation}}

\def\beqa{\begin{eqnarray}}
\def\eeqa#1{\label{#1}\end{eqnarray}}
\def\eeqan{\end{eqnarray}}

\let\bar=\overbar

\def\Dslash{\not{\hbox{\kern-4pt $D$}}}
\def\dslash{\not{\hbox{\kern-2pt $\del$}}}

\def\msb{{\bar{\ssstyle M \kern -1pt S}}}